\documentstyle[fleqn]{article}
\title{\bf Non-universal dynamics of staggered non-equilibrium particle
systems and Ising chains}
\author{R. B. Stinchcombe$^1$, J. E. Santos$^1$ and M. D. Grynberg$^2$\\
1-{\it Department of Physics - University of Oxford}\\
{\it Theoretical Physics, 1 Keble Road,
 Oxford OX1 3NP, UK}\\
2-{\it Departamento de F\'{\i}sica, Universidad Nacional de La Plata.}\\
{\it (1900) La Plata, Argentina}\\
 PACS numbers: 05.70.Ln, 75.10.Hk, 02.50.Ey\\
\\
Published in J. Phys. A {\bf 31}, 541--549 (1998).}
\date{}
\newcommand{\beq}{\begin{equation}}
\newcommand{\eeq}{\end{equation}}
\newcommand{\ket}[1]{\mbox{$ \mid #1\, \rangle$}}
\newcommand{\bra}[1]{\mbox{$ \langle\, #1\mid$}}

\newcommand{\spx}[1]{\mbox{$\hat{\sigma}^{x}_{#1}$}}
\newcommand{\spz}[1]{\mbox{$\hat{\sigma}^{z}_{#1}$}}

\newcommand{\undn}{\mbox{$\underline{n}$}}

\begin{document}
\maketitle
\newpage
\begin{abstract}
 Non-universal dynamics is shown to occur in
 a one-dimensional non-equilibrium system of hard-core
 particles. The stochastic processes included are pair
 creation and annihilation (with rates $\epsilon$ and $\epsilon'$)
 and symmetric hopping rates which alternate from one bond to
 the next ($p_A$, $p_B$). A dynamical scaling relation between the
 relaxation time and the correlation length in the steady state is
 derived in a simple way for the case
 $\epsilon'>p_A\gg p_B\gg\epsilon$. We find that the
 dynamical exponent takes the non-universal value
 $z=2\ln(\frac{\epsilon'}{\epsilon})/
\ln(\frac{p_B\epsilon'}{p_A\epsilon})$.

 For the special condition $\epsilon+\epsilon'=
 p_A+p_B$, where the stochastic system is in principle
 soluble by reduction to a free fermion system, the model
 is mapped to the Glauber dynamics of an Ising chain
 with alternating ferromagnetic bonds
 of values $J_1$ and $J_2$, in contact with a
 quantum thermal bath. The full time
 dependence of the space-dependent magnetisation
 and of the equal time spin-spin correlation function
 are studied by  writing the master equation
 for this system in the quantum Hamiltonian
 formalism.
 In particular we obtain the dispersion
 relations and rigorously confirm the results
 obtained for the correlation length
 and for the dynamical exponent.
 \end{abstract}
\pagebreak
\section{Introduction}
 This paper discusses non-universal critical
 dynamics of non-equilibrium particle systems, in
 which dynamic exponents depend on microscopic parameters
 (ratios of transition rates).

 Non-equilibrium particle systems with stochastic dynamics
 show properties ranging from steady state phase transitions
 to turbulence, shocks, and non-equilibrium analogues of
 strong fluctuations and critical behaviour \cite{Derrida,Schmittman}.
 As with equilibrium
 critical phenomena, associated critical exponents are typically
 dependent on symmetries and dimensionalities and (for dynamics)
 conservation laws, but not on microscopic details. In particular
 the dynamic exponent $z$ in many simple systems
 describes diffusive relaxation towards equilibrium ($z=2$),
 while in others it takes other parameter independent
 values, e.g. $z=3/2$ in the one-dimensional noisy KPZ or Burgers
 equations \cite{Kardar,Forster,Frey1,Frey2}.

 Simple hard-core particle models of non-equilibrium stochastic
 dynamics can be mapped to quantum systems and in particular to
 spin models. This latter mapping is achieved either by a pseudo-spin
 representation of the particles or by associating particles with
 domain walls \cite{Alcaraz,Gwa,Barma,Family,Grynberg}.

 But it is well known that non-universal behaviour
 can exist in certain non-uniform spin systems with
 Glauber dynamics \cite{Glauber}, such as that providing a simple model
 for freezing into non-equilibrium states and glassy dynamics
 \cite{Cornell1,Cornell2,Cornell3,Jackle}.
 Here the relationship $\tau\sim\xi^z$
 between the relaxation time $\tau$ and the equilibrium
 correlation length involves a parameter-dependent dynamic
 exponent $z$ \cite {Droz,Deker,Haake}.
 This suggests that analogous non-universal behaviour can
 exist in non-uniform stochastic particle systems, and this
 is demonstrated here.

 The model considered is defined on a one-dimensional lattice
 with $L$ sites. We divide this into two sub-lattices, $A$ and $B$,
 containing even and odd sites respectively. Each site can
 be occupied by at most one particle.
 A pair of particles in adjacent sites can annihilate with rate
 $\epsilon'$. A pair can be created on two empty adjacent sites
 with rate $\epsilon$. A particle can hop to an empty nearest
 neighbour site with rate $p_{A}$ if it is in sub-lattice $A$ and
 with rate $p_{B}$ if it is in sub-lattice $B$.

 The correlation lengths characterising the steady state particle
 separations on each sub-lattice can be obtained by dynamic balance
 conditions within a mean field approximation. The characteristic
 time for diffusion across the smaller of the lengths, $\xi$, can then
 be estimated from the hopping process. This is particularly simple
 to do when e.g. the annihilation rate is much greater than the creation
 rate (when $\xi$ is large) and one hopping rate is much greater than
 the other. This simple procedure already provides the non-universal
 behaviour of the dynamic exponent $z$.

 For a special case where the rates satisfy
 one constraint, the quantum spin
 Hamiltonian representing the particle
 dynamics reduces to a free fermion
 form after a Jordan-Wigner transformation
 \cite{Jordan}. The alternative and
 simpler way used here of achieving
 an exact solution for this case is
 by mapping the stochastic particle
 system to the Glauber dynamics of an
 Ising model with alternating bonds
 $J_1$, $J_2$ \cite{Family,Racz}.

 The correlation lengths and relaxation time
 readily provided by the simple
 argument in the particle picture thus translate, for
 the special rate relation, into results for the alternating bond
 Ising model, where they can be confirmed by exact
 calculation on the spin model.

 The Glauber dynamics of this model is that resulting from
 coupling the static Ising system to a quantum thermal bath
 and considering the Van Hove limit of a weak interaction and
 very large times.
 By writing the master equation which describes the Glauber
 dynamics in a quantum Hamiltonian formalism,
 we can easily obtain the
 equation determining the time evolution of the equal-time
 spin-spin correlation function.
 From this quantity one can compute the equilibrium domain wall
 density in the Glauber
 problem which is equivalent to the steady state density
 of particles in the stochastic particle dynamics
 \cite{Family,Racz}.
 By taking the Fourier transform of this equation
 we obtain a system of four linear equations
 which  determine the dependence of the
 eigen-frequencies with the wave vectors of the
 two spin excitations. The solution of the
 associated secular equation gives us the
 dispersion relation.
 The behaviour of the frequency at low
 wave vectors determines the
 critical exponent $z$. We
 fully confirm the results obtained previously.

 The structure of this paper is as follows:
 in section 2, we define the dynamics of
 the particle system in terms of the
 constituent stochastic processes
 (diffusion, annihilation and creation).
 Using the simple mean-field
 and random-walk arguments referred to above
 we obtain the relation
 between the relaxation time and
 the equilibrium correlation
 length and hence we determine the
 critical exponent $z$.
 In section 3 we present the Glauber dynamics
 of the alternating-bond Ising system,
 giving the rates of transition between the
 different configurations.
 We then go on to relate the spin flip
 processes to domain wall hopping,
 creation and annihilation, which
 translate to the particle processes.
 In section 4, we write down the master equation
 in a quantum Hamiltonian formalism and
 derive the equation giving the time evolution of
 the spin-spin correlation function. Hence,
 we obtain the dispersion relation and
 from it we can again compute $z$.
 Finally, in section 5 we present our
 conclusions.

 \section{The reaction-diffusion system and its non-universal
 dynamics}

 The hard-core particle dynamics is as specified in the
 Introduction: a particle on the even sub-lattice may hop to its left
 nearest neighbour site, provided it is empty, or to its
 right nearest neighbour (if empty) at rate $p_A$; the
 corresponding rate for hopping from the odd sub-lattice is
 $p_B$; in addition, pairs of particles can appear at
 (or annihilate from) adjacent empty (or full) sites at
 rate $\epsilon$ (or $\epsilon'$). These processes are
 depicted in Table 1 (see below).

 We now consider the description of the steady state.
 Here the particle densities $\rho_A$ and $\rho_B$ on
 the two sub-lattices will be uniform.
 Since the system is in equilibrium the processes of creation
 and annihilation have to balance each other. So one concludes
 that provided the mean field approximation applies,
 $\epsilon'\,\rho_A\rho_B=\epsilon\,(1-\rho_A)(1-\rho_B)$.
 The same must be true for the processes of diffusion.
 Hence $p_A\,\rho_A (1-\rho_B)=p_B\,(1-\rho_A)\rho_B$. One obtains
 from these relations $\rho_A\sim\frac{x_A}{1+x_A}$, where
 $x_A=(\frac{\epsilon\,p_B}{\epsilon'\,p_A})^{1/2}$ and
 $\rho_B\sim\frac{x_B}{1+x_B}$, where
 $x_B=(\frac{\epsilon\,p_A}{\epsilon'\,p_B})^{1/2}$. These
 densities allow us to extract two distinct separation
 lengths, the average distances $\rho_{\alpha}^{-1}$
 between particles on each sub-lattice $\alpha$.

 We will be particularly interested in the critical
 situation where these two characteristic lengths are both
 large. That occurs when $\epsilon'\gg\epsilon$, giving large
 particle separations, or where $\epsilon\gg\epsilon'$
 (large vacancies separations). For convenience we consider
 only the former case. It is also convenient to take $p_A\gg p_B$,
 to widely separate the two lengths. Both are still large provided
 $\epsilon'/\epsilon\,\gg\,p_A/p_B$. The shorter, controlling correlation
 length $\xi$ is then
 \beq
 \xi\;\sim\;\rho_B^{-1}\;\sim\;x_B^{-1}
 \;=\;(\epsilon'p_B/\epsilon p_A)^{1/2}\enspace.
 \label{eq1}
 \eeq

 Now to determine the rate of
 approach to this steady state
 for $\epsilon'\gg \epsilon$ one uses
 a modification, appropriate to the particle
 dynamics, of an argument given by \cite{Cordery}.
 For large times the evolution of the system
 is determined by the limiting relaxational
 process, namely particles
 diffusing until they meet, when they annihilate.
 For situations close to the steady state, i.e. for
 long time critical dynamics (starting from generic
 non-equilibrium initial states), the characteristic time
 involved is that for particles to diffuse across the
 shorter steady state separation length, $\xi$,
 and to annihilate. Now to traverse
 two adjacent bonds both rates $p_A$ and $p_B$ enter and the
 effective diffusion rate is $\frac{4\,p_A\,p_B}{p_A+p_B}$.
 Thus the characteristic time is $\tau\sim(\epsilon'p_B/\epsilon p_A)\,
 \frac{p_A+p_B}{4p_A p_B}\propto\xi^z$ giving
 \beq
 z=2\ln(\frac{\epsilon'}{\epsilon})/
 \ln(\frac{\epsilon'p_B}{\epsilon p_A})
 \label{eq2}
 \eeq
 in the regime of validity ($\epsilon'/\epsilon\gg p_A/p_B\gg 1$).
 This is a non-universal result, being dependent on ratios of rates.
 The above reasoning is based
 on the mean field argument that we
 have used to determine $\rho_A$ and $\rho_B$.
 Mean field theory determines correctly
 the characteristics of the steady state
 but it fails to predict the correct approach to the
 steady state \cite{Grynberg}.
 This is due to the diffusion aspects
 of the problem, which are relevant in one dimension.
 The random-walk argument that we have given to
 determine $\tau$ incorporates these aspects.
 We will see in the next sections that the
 same results are obtained by an exact calculation
 based on an equivalent Glauber-Ising dynamics. That
 equivalence applies when the rates satisfy
 \beq
 \epsilon+\epsilon'\;=\;p_A+p_B\enspace.
 \label{eq3}
 \eeq
 This relation is also sufficient to reduce
 the quantum Hamiltonian representing the particle
 dynamics to a free fermion form, which
 gives another possible
 method for solving the particle dynamics exactly.

 Finally, we note that the exponent $z$ changes continuously with
 the value of the rates  of the diffusion and reaction
 processes, as given by equation (\ref{eq2}). 
 A similar example of a critical exponent which changes continuously
 was found in the study of the dynamics of the $q$-state Potts model 
 \cite{Derr96}. In this system, the fraction of spins which
 never flip up to time $t$,  $r(q,t)$, decays like a power law $r(q,t)
 \sim t^{-\theta(q)}$ when the initial condition is random.                
 The exponent $\theta(q)$ varies continuously with $q$.

\section{An Ising system with generalised Glauber
 dynamics, and its relationship to reaction-diffusion processes}
 The Ising system considered is a spin $\frac{1}{2}$ chain, where
 a spin on the even
 sub-lattice is coupled to its left nearest neighbour
 by a ferromagnetic Ising interaction of strength $J_1$ and
 coupled to its right nearest neighbour by an interaction
 of strength $J_2$ ($J_1>J_2$) \cite{Droz}.
 One can therefore write
 the Hamiltonian for this alternating-bond system in the form
\beq
 H\;=\;-\sum_{l=1}^{L}\,J_{l}\sigma_{l}\sigma_{l+1}
 \label{eq4}
\eeq
 where $\sigma_{l}$ denotes the
 eigenvalues of $\spz{l}$, $J_{l}=J_1$
 for $l$ odd and $J_{l}=J_2$
 for $l$ even. The transitions between
 different configurations of the system, i.e. between
 different sets of values of the spins, are of the
 Glauber type, i.e. the system evolves by single spin
 flips  \cite{Glauber}. The probability per unit time
 that a spin will flip from its value to the
 opposite one is taken to be \cite{Droz}
\beq
\omega(\sigma_{l})\;=\;\frac{\Gamma}{2}\,
 [\,1-\frac{1}{2}\,\sigma_{l}(\gamma_l^{+}\sigma_{l-1}\,+\,
 \gamma_l^{-}\sigma_{l+1})\,]
\label{eq5}
\eeq
 with
\beq
\gamma_{l}^{\pm}\;=\;\tanh(K_{l-1}+K_l)\,\pm\,\tanh(K_{l-1}-K_{l})
\label{eq5A}
\eeq
 where $K_l=J_l/\mbox{$k_B$T}$. Notice that if we take
 $J_1=J_2$ we will recover the usual Glauber rates.
 It can be easily verified that these rates satisfy
 detailed balance which is a sufficient condition for the
 steady state distribution of the Ising system to be a Gibbs
 distribution.
 This dynamics can be derived \cite{Martin} by coupling the Ising
 system, described by Hamiltonian (\ref{eq4}) to
 an ensemble of free fermion baths with a grand canonical probability
 distribution, i.e. with a density matrix given
 (in the case of no interaction) by
 $\hat{\rho}\;=\;\exp[-\beta(\hat{H}-\mu\hat{N})]/Z$
 where $\hat{H}$ and $\hat{N}$ are respectively, the
 Hamiltonian and the particle number operator of the
 free fermion bath, $\beta$ is the inverse temperature
 and $\mu$ the chemical potential. The quantity
 $Z$ is the grand partition function and is just a normalisation
 factor. If the particle number of the fermion system is kept fixed,
 then the chemical potential is, for low enough temperatures,
 essentially equal to the Fermi energy of the system.
 The coupling is done via an interaction operator which
 couples the $\spx{}$ component of each spin to a
 thermal bath whose probability distribution is the one
 given above. The different copies of the thermal bath
 are totally uncorrelated.
 If one takes the joint limit of the interaction
 strength $\lambda$ going to
 zero and  $t\rightarrow\infty$ such that $t\lambda^{2}$
 is a constant (limit of Van-Hove) \cite{Hove1,Hove2}
 one obtains the transition
 rates given by (\ref{eq5}), in the condition that the chemical
 potential (Fermi energy) of the fermion bath is much larger
 than the couplings $J_1$ and $J_2$. If that is not the case
 then the transition rates still have the form (\ref{eq5}), but
 the parameter $\Gamma$ is no longer a constant and depends on
 the value of the neighbouring spins. We will consider that
 we can take $\Gamma$ as a constant.

 It is well known that there is a duality relation between
 Glauber dynamics and the reaction-diffusion model of hard-core
 particles discussed above \cite{Family,Racz}. To see this
 we take a given configuration
 of the Ising system and consider the lattice of sites located
 in the middle of the bonds between the Ising spins (dual lattice).
 If the neighbouring Ising spins have different signs then we place
 a particle at that site of the dual lattice. Otherwise we leave the
 site empty. In this way we map domain walls in the Ising system to
 particles in the dual lattice. It can be shown that this
 mapping has a precise mathematical meaning \cite{Santos}.
 The different possible
 processes of transition for a given spin, its translation in
 terms of particle processes, the rates associated and
 their relation are given in table 1.
\begin{table}[t]
\begin{center}
\large
\begin{tabular}{|c|c|c|} \hline
 Initial State & Final State  & Transition Rate\\ \hline
 $\uparrow\circ\uparrow\circ\uparrow$ &
$\uparrow\bullet\downarrow\bullet\uparrow$
& $\frac{\Gamma}{2}\,[1-\tanh(K_1+K_2)]=\epsilon$ \\ \hline
$\uparrow\bullet\downarrow\bullet\uparrow$&
$\uparrow\circ\uparrow\circ\uparrow$
& $\frac{\Gamma}{2}\,[1+\tanh(K_1+K_2)]=\epsilon'$\\ \hline
$\uparrow\circ\uparrow\bullet\downarrow$&
$\uparrow\bullet\downarrow\circ\downarrow$
& $\frac{\Gamma}{2}\,[1+\tanh(K_1-K_2)]=p_A$\\ \hline
$\uparrow\bullet\downarrow\circ\downarrow$&
$\uparrow\circ\uparrow\bullet\downarrow$&
$\frac{\Gamma}{2}\,[1-\tanh(K_1-K_2)]=p_B$\\ \hline
\end{tabular}
\caption{Processes of transition for a spin in the
 even sub-lattice and the equivalent particle
 processes. The spin which is to flip is the
 central one. The coupling strength of the left bond is $J_1$,
 and of right bond $J_2$. An empty circle in the
 dual lattice is to be identified with a vacancy and a full
 circle with the presence of a particle. 
 For a spin in the odd sub-lattice the rates of the last 
 two processes should be interchanged.}
\end{center}
\end{table}
 Notice that the sum of the rates of the first two processes
 is equal to the sum of the last two in agreement with (\ref{eq3}).
 This shows that this system
 is in the class of systems that are integrable (the precise
 meaning of this word will be made clear below) through
 free fermions \cite{Alcaraz}.
 This allows the computation of correlation
 functions for a set of distributions of
 initial configurations
 \cite{Grynberg,Santos}. Here we will not pursue this point.
 If we now take the limit of low temperatures ($K_1\gg K_2\gg 1$)
 we see that the rates of pair annihilation  ($\epsilon'$)
 and of diffusion
 from the strong bond to the weak bond $p_A$ are equal to
 $\Gamma$. On the other hand the rate of diffusion from
 a weak bond to a strong bond is $p_B\sim\Gamma\exp(-2(K_1-K_2))$.
 Finally the rate of pair creation
 is $\epsilon\sim\Gamma\exp(-2(K_1+K_2))$.
 So we see that these two last processes are exponentially suppressed at
 low temperatures. Nevertheless they have to be taken into account for
 a proper description of the steady state and diffusive dynamics,
 as we saw in section 2. We will see in the next section that the results
 given in section 2 may be obtained by an exact calculation.

\section{The master equation in the quantum Hamiltonian
 formalism}

In this section we will study the time evolution
of the spin-spin correlation function. For that we need the full
master equation for the evolution of the Ising system with
rates given by (\ref{eq5}). Since the master equation is a linear
equation a particularly convenient way to write it is to use
an operator formalism which assigns to each configuration
of Ising spins a vector $\ket{\undn}$ in an Hilbert space.
The probability distribution for the different configurations
at a given time $t$ can then be written as a state-vector
$\ket{\Psi_t}\;=\;\sum_{\undn}\,P(\undn,t)\,
\ket{\undn}$ \cite{Kadanoff}, where $P(\undn,t)$
is the probability to find  configuration $\undn$
at time $t$ and is a solution of the master equation.
The set
of different $\ket{\undn}$  is supposed 
to be orthonormal and complete.
One can then  write the master equation in the
compact form $\partial_{t}\ket{\Psi_t}=-\,\hat{T}\,\ket{\Psi_t}$,
where $\hat{T}$ is a linear and in general non-hermitian operator
that for two state systems with local interactions can be written
as a quantum spin Hamiltonian. The average values of quantities
like the Ising spins can also be conveniently
represented in this language as $\langle\sigma_j(t)\rangle
\;=\;\bra{s}\,\spz{j}\,\ket{\Psi_t}$, where
$\spz{j}$ is the Pauli spin matrix at site $j$
and $\bra{s}\;=\;\sum_{\undn}\,\bra{\undn}$. Substituting
the state vectors and the operators by their definitions
and using the orthonormality relations one sees
that one obtains the usual definition of average over configurations.
One also can, given the correspondence with the Schr\"{o}dinger
equation, define an Heisenberg representation of
the operators by $\hat{A}(t)\;=\;e^{\hat{T}t}\,\hat{A}\,e^{-\hat{T}t}$,
where $\hat{A}$ is a generic operator. These operators
obey the equation of motion $\frac{d\hat{A}}{dt}\;=\;
[\,\hat{T}\,,\,\hat{A}(t)\,]$. It should be stressed that
this is only a convenient way to represent the master
equation for this system and is not related to the
intrinsic quantum dynamics of the Ising system which is
given by the Hamiltonian (\ref{eq4}). In fact, the dynamics
is in this case generated by the interaction of the Ising
spins with the thermal bath and cannot be deduced from
the form of (\ref{eq4}) alone.
For the system that
we have studied in the previous
section the operator $\hat{T}$
has the form
\beq
\hat{T}\;=\;\frac{\Gamma}{2}\,\sum_{l=1}^{L}\,(1-\spx{l})\,
[\,1-\frac{1}{2}\,\spz{l}(\gamma_l^{+}\spz{l-1}\,+\,
 \gamma_l^{-}\spz{l+1})\,]
\label{eq6}
\eeq
as can be seen if we consider the matrix elements
$\bra{\undn'}\,\hat{T}\,\ket{\undn}$ and
$\bra{\undn}\,\hat{T}\,\ket{\undn}$, where $\undn'$ is
a configuration differing from $\undn$ by the flip of a
single spin $\sigma_{l}$, say. We obtain
\begin{eqnarray}
\bra{\undn'}\,\hat{T}\,\ket{\undn}&=&-\frac{\Gamma}{2}\,[
\,1\,-\,\frac{1}{2}\sigma_{l}\,
(\gamma_{l-1}^{+}\sigma_{l-1}\,+\,
\gamma_{l+1}^{-}\sigma_{l+1})\,]\nonumber \\
\bra{\undn}\,\hat{T}\,\ket{\undn}&=&
\frac{\Gamma}{2}\,\sum_{l=1}^{L}
[\,1\,-\,\frac{1}{2}\sigma_{l}\,
(\gamma_{l-1}^{+}\sigma_{l-1}\,+\,
\gamma_{l+1}^{-}\sigma_{l+1})\,]
\label{eq7}
\end{eqnarray}
where $\sigma_{l}$, etc., are the values of the Ising spins
in configuration $\undn$.
The first relation gives (up to a minus sign) the transition rate
between configurations $\undn$ and $\undn'$. The second relation
gives the total rate of transition out of configuration $\undn$. This
agrees with the general definition of the $\hat{T}$ operator
\cite{Feld}.
The form (\ref{eq6}) and associated equations
of motion are particularly convenient
for they allow the calculation of multiple time spin-spin
correlation functions. In the same way one can represent the
master equation for the particle dynamics in terms of a $\hat{T}$
operator which is also a quantum spin Hamiltonian.
The duality relation between these two models can
be given a precise mathematical meaning by means of a
similarity transformation between the operators
of the two models \cite{Santos}. In particular, the
$\hat{T}$ operator of the generalised Glauber
dynamics maps to the $\hat{T}$ operator of the
particle system. This later operator can be
expressed in terms of free fermions by a Jordan-Wigner
transformation \cite{Jordan} and therefore
the system is completely integrable.
Many correlations functions relevant for the
study of the particle dynamics and the associate
Glauber problem
can hence be obtained.

If  one now considers the Heisenberg equation of motion
for $\spz{j}(t)$ and takes the average value
on some arbitrary initial state
$\ket{\Psi}$ in the way indicated above
one obtains the differential
equation giving the time development of the average
space dependent magnetisation \cite{Droz}.
This equation can be solved
by defining the Fourier transforms $m_{k}^{+}$ and $m_{k}^{-}$ of
the magnetisation at even and odd sites respectively, with
the wave vector $k$ equal to
$k=\frac{2\pi}{L}n$ with $n$ an integer and such that $k$ lies
in the limits
$-\frac{\pi}{2}\leq k<\frac{\pi}{2}$  (we have chosen the lattice
spacing to be one).
Notice that the Brillouin zone has been reduced to half
of the size it would have if $J_1=J_2$ since the lattice has now
a periodicity of two.
This procedure gives us a system of two equations
coupling $m_{k}^{+}$
and $m_{k}^{-}$ for every $k$. The relaxational eigenvalues
of this system are
\beq
\omega_{k}^{\pm}\;=\;1\,\pm\,[
\cos^{2}k\,\tanh^{2}(K_1+K_2)\,+\,
\sin^{2}k\,\tanh^{2}(K_1-K_2)]^{1/2}
\label{eq9}
\eeq
where $\omega_k^{\pm}$ stands for the lower (acoustical)
and upper (optical) branches of this dispersion relation.
The late (critical) dynamics is determined by the low $k$
 modes of the acoustical branch $\omega_k^-$.
The characteristic length associated with the decay
 of the local magnetisation in the steady  state
 can be found from the value of $\mid k \mid$ ($=\frac{2\pi}{\xi}$)
 making the right hand side of (\ref{eq9}) vanish. We therefore
 take the analytic continuation of (\ref{eq9}) to the complex plane.
 We obtain  the space decaying mode with infinite lifetime (which
 decays over a characteristic length $\xi$). For low temperatures
 $\xi$ is given by
\beq
\xi\;\sim\;\exp (2 K_2)\enspace.
\label{eq10}
\eeq
Therefore at low $k$ and low temperatures one can write
the dispersion relation as
\beq
\omega_{k}^{-}\;=\;2\Gamma\xi^{-z}\,(1+(k\xi)^{2})
\label{eq11}
\eeq
with $\xi$ given by (\ref{eq10}) and
\beq
 z=2+\frac{J_1-J_2}{J_2}
\label{eq12}
\eeq
a result given in \cite{Droz}. Now to determine
the time evolution of the density of particles we
need to compute the equivalent quantity in the Glauber
language. This quantity is the domain wall local density
\cite{Family} which is given by
$\frac{1}{2}(1-\langle\sigma_l(t)\sigma_{l+1}(t)\rangle)$.
We therefore need the equation giving the time evolution
of the spin-spin correlation function. 
Notice that these correlations do not involve only eigenmodes
of (\ref{eq6}) orthogonal to the slowly decaying ones, so their relaxation
follows closely the slowest processes in the system.
Using the transfer matrix formulation we can write
$\frac{d}{dt}\bra{s}\spz{l}(t)\spz{m}(t)\ket{\Psi}\;=\;
\bra{s}[\,\hat{T}\,,\,\spz{l}(t)\spz{m}(t)\,]\ket{\Psi}$
where $\ket{\Psi}$ is as above a general initial state.
Substituting $\hat{T}$ by its expression given in
(\ref{eq6}) we obtain the equations
\begin{eqnarray}
\frac{d}{dt}\langle\sigma_l(t)\sigma_{m}(t)\rangle\!\!&=&\!\!
-2\Gamma\langle\sigma_l(t)\sigma_{m}(t)\rangle\,+\,
\frac{\Gamma}{2}[\,\gamma_l^{+}\langle\sigma_{l-1}(t)\sigma_{m}(t)\rangle
\,+\,
\gamma_l^{-}\langle\sigma_{l+1}(t)\sigma_{m}(t)\rangle
\nonumber\\
& &\mbox{}+\gamma_m^{+}\langle\sigma_{l}(t)\sigma_{m-1}(t)\rangle
\,+\,
\gamma_m^{-}\langle\sigma_{l}(t)\sigma_{m+1}(t)\rangle\,]
\label{eq13}
\end{eqnarray}
if $l\neq m$ and the boundary condition
$\langle\sigma_l(t)\sigma_{l}(t)\rangle\,=\,1$.
Since these equations are linear this boundary condition
may be ignored. It can be later enforced by superposing
different solutions \cite{Glauber}. To solve these equations
we introduce  Fourier transforms of the quantities
$\langle\sigma_l(t)\sigma_{m}(t)\rangle$ when $l$ and $m$ are
even, $l$ is even and $m$ is odd, etc. The two wave vectors
$k$ and $k'$ which now appear are defined as above. We obtain
a system of four linear equations. The associated secular equation
has four solutions corresponding to having two excitations in
the acoustic branch, one in the acoustic and one in the optical
branch (two
solutions) and two excitations in the optical branch. The frequency
$\omega_{kk'}$ is given by
\beq
\omega_{kk'}\;=\;\omega_k+\omega_{k'}
\label{eq14}
\eeq
 where $\omega_k$ and $\omega_{k'}$ are given by one of the solutions in
(\ref{eq9}).
If we now take the translational average of
the correlation function
$\langle\sigma_l(t)\sigma_{l+1}(t)\rangle$ we
see that only the terms for which $k'=-k$ remain
in the Fourier sum. Hence the decay of density of domain walls
(density of particles) is determined by the
acoustical branch of  $\omega_{k,-k}$
for low $k$. At low temperatures we just obtain
$\omega_{k,-k}=2\omega_k^{-}$ with $\omega_k^-$ given
(\ref{eq11}). The interparticle distance in the
steady state is given by the value of $k$ that makes
$\omega_{k,-k}$ vanish. This coincides with (\ref{eq10}).
The relaxation time $\tau$ is  given by $\tau=\xi^{z}/4\Gamma$
($\tau^{-1}=\omega_{0,0}$). This translates into (\ref{eq2}) using the
relationships between rates in the Ising and particle
pictures (see Table 1) in the low temperature limit.
So we recover the results obtained in section 2
in a rigorous way.
It can also be seen from  (\ref{eq14})
that dynamical
scaling still holds for this model.
This is nontrivial,
since as pointed out above there are
two correlation lengths in the
problem.
\section{Conclusions}

We have studied a generalised reaction-diffusion process, and the
related model of generalized Glauber dynamics for a staggered
Ising model. We obtained the following results

i) The dynamical exponent $z$ relating the relaxation time to
the equilibrium correlation length is non-universal, depending
on the ratios of rates or the ratio between the
couplings of the Ising model.

ii) We found that dynamical scaling still holds, i.e. the
dispersion relation still depends on a function of a single
parameter $k\xi$ where $k$ is the wave-vector and $\xi$ is
the equilibrium correlation length.

It would be interesting to investigate the exact solution
of this problem from the point
of view of the particle dynamics, since one
can diagonalize
the dynamical operator $\hat{T}$ by using free fermions.
Also it would be interesting to study if there is any change
in the dynamical exponent $z$ when one goes away from the
free fermion condition. One would expect, based on the general
argument given in section 2 that the relationship of $z$ to the
 basic rates does not
change provided that $\epsilon'/\epsilon\gg p_A/p_B \gg 1$.
This means that in the long-time limit the system has a very
low density of particles and therefore one should expect that
the interactions are unimportant. Preliminary simulations seem
to confirm this result \cite{Grynberg2}.
\\ \\
{\noindent
{\bf Acknowledgments:}
 It is a pleasure to acknowledge many
 helpful discussions with Gunter Sch\"{u}tz.
 We would also like to thank the International
 Center of Theoretical Physics for kind hospitality
 during the early stages of this work. R.B.S.
 would like to thank Rutgers University for
 kind hospitality during the later stages of this
 work. J. E. S. is  supported by the
 Grant: PRAXIS XXI/BD$/3733/94$ - JNICT -
 PORTUGAL.}
\pagebreak

\end{document}